\documentclass[fleqn,12pt,twoside]{article}
\usepackage{espcrc1}

\usepackage{epsfig}

\newcommand{\AmS}{{\protect\the\textfont2
  A\kern-.1667em\lower.5ex\hbox{M}\kern-.125emS}}

\title{Deep Exclusive Scattering and Generalized Parton Distributions: Experimental Review}

\author{F.~Sabati\'e\address[Saclay]{CEA-Saclay, Service de Physique Nucl\'eaire, F91191 Gif-sur-Yvette, France\\
Email: fsabatie@cea.fr}}
       
\begin{document}

\maketitle

\begin{abstract}
Since the Generalized Parton Distribution theoretical framework was introduced in the late 90's, a few published
and numerous preliminary results from Deep Exclusive Scattering (DES) have been extracted from non-dedicated experiments
at HERA and Jefferson Lab.
We review most of these results, comment on the ongoing dedicated research in this topic and conclude with the
expectations from the next generation of experiments in the near future.
\end{abstract}

\section{INTRODUCTION: DEEP EXCLUSIVE SCATTERING (DES)}

Deep Exclusive Scattering reactions are special cases of hard scattering processes, which play a very
important role in the understanding of the quark and gluon structure of the nucleon. The key is that the
hard probe selects a small configuration, be it quark, antiquark or gluon inside the nucleon, which
interaction can be described by means of perturbation theory. The non-perturbative part consists in the
remaining system and can be treated separately. This separation of a perturbative and non-perturbative
part in hard scattering, also known as factorization \cite{fact}, has been extensively used to gain information about the
non-perturbative structure of the nucleon. For instance in Deep Inelastic Scattering (DIS), it led to
the discovery of the quark and gluon substructure of the nucleon, and later, to the measurement of the momentum and
spin distribution of quarks and gluons inside the nucleon.

Nowadays, with the advent of moderate to high energy and high luminosity lepton accelerators coupled with high
resolution and/or large acceptance detectors, it is possible to extend the landscape of hard scattering from
inclusive processes such as DIS to exclusive processes, which contain a wealth of new information about the
nucleon structure. Similar to parton distributions for DIS, the Generalized Parton Distributions which can be
obtained through the study of Deep Exclusive Scattering processes contain information on the quark/antiquark
and gluon correlations, and more specifically on both the transverse spatial and longitudinal momentum dependences.
The pioneering articles on the subject can be found in \cite{gpd1,gpd2,gpd3}.

The simplest Deep Exclusive Scattering process is called Deeply Virtual Compton Scattering (DVCS), and consists in the 
lepto-production of real photons from the proton in the Bjorken regime (large $Q^2$ and large photon energy $\nu$ at fixed $x_{B}$).
It is considered the simplest since the outgoing real photon has no structure, whereas in the lepto-production of mesons
for instance, both the structure of the meson and the nucleon can be considered as unknown. These proceedings will focus on
DVCS but more information on meson lepto-production can be found elsewhere \cite{dvmp_hermes,dvmp_clas}. It is no surprise that most
published work about GPDs focus on DVCS as described later in this article. Not only is DVCS the simplest process, but it has
also a very interesting feature: the real photon can either be emitted from one of the lepton lines (Bethe-Heitler process, BH)
or from the nucleon (DVCS). Both processes are not distinguishable, therefore they interfere at the amplitude level. By measuring
the difference of cross-sections for opposite lepton helicities, one is sensitive to the interference term alone, which
is basically the product of the DVCS amplitude imaginary part by the BH amplitude\footnote{This is an easily calculable
process using QED and the knowledge of form factors at low $t$.}. This is especially useful at moderate energy at Jefferson Lab
or HERMES since Bethe-Heitler is strong in these kinematics and the cross-section difference is therefore sizeable. The only
difficulty of such measurements lies in the need for exclusivity, which usually means the detection of the three-particle final state.
This is what lacks in all re-analysis of old data and it is precisely what the new generation of experiments will bring,
with dedicated setups.

The following section will first examine published data of experiments at HERA and Jefferson Lab. Three experiments
running at Jefferson Lab with dedicated experimental setup will then be described. Finally, the rest of the article will make
a summary of ambitious projects at H1 and HERMES at HERA, COMPASS at CERN and Jefferson Lab,
which will eventually give strong constraints on GPDs in a wide kinematical domain.

\section{PUBLISHED DATA ON DVCS}
\label{oldies}

After theorists pointed out the huge interest in making DVCS measurements in various kinematical domains, physicists from
HERA and Jefferson Lab started re-analyzing data to look for photon lepto-production events. All experiments
succeeded in extracting a signal from data even though they suffered from two obvious flaws: a far-from-ideal exclusivity
since in all cases, only two out of three particles were detected in the final state; a rather limited statistics. However,
these studies have paved the way to improved experimental setup in their respective facility, and demonstrated that such
experiments are now doable.

\subsection{DVCS measurements at high energy}

\begin{figure}[hbt]
\begin{center}
\epsfig{file=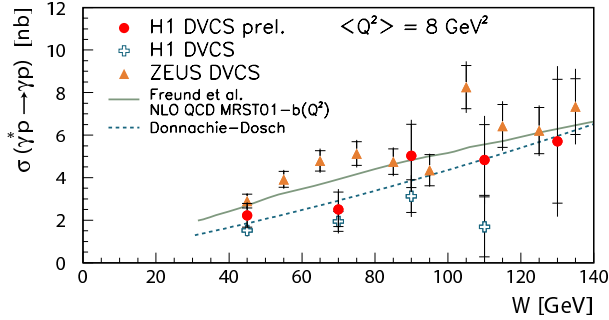,width=0.5\linewidth}
\end{center}
\caption{Reduced DVCS cross-section as a function of $W$ for $<Q^2>=8$~GeV$^2$. Published H1 \cite{dvcs_h1}, preliminary H1 and
published ZEUS data \cite{dvcs_zeus} are represented. The measurement is in fair agreement with a NLO QCD prediction using GPD parametrisation.}
\label{fig1}
\end{figure}

In a very high energy regime where gluon contributions are dominant,
H1 and ZEUS at HERA were able to extract reduced DVCS cross-sections from data,
using events in which they detected the scattered positron and outgoing real photon \cite{dvcs_zeus,dvcs_h1}.
The proton remained undetected since
the expected transfer is small and the proton therefore escapes at small angles. The data in both cases were integrated
along the azimuthal angle $\varphi$, which is the angle between the leptonic and hadronic planes. This integration cancels
most of the interference term, allowing for a subtraction of the Bethe-Heitler contribution, which is calculable. Figure~\ref{fig1}
shows the published H1 and ZEUS data along with the preliminary H1 data. The reduced cross-section is measured in an
extended kinematical regime, $4<Q^2<80$~GeV$^2$, $40<W<140$~GeV and $|t|<1$~GeV$^2$. All data samples are in reasonable
agreement. The cross-section was compared to NLO QCD predictions using GPD parametrisations which reproduced the
data within error bars. 

%

\subsection{DVCS measurements at moderate energy}

At lower energy, the Bethe-Heitler dominates the cross-section. As explained in the introduction, it is still possible to
make a meaningful measurement using a polarized lepton beam and measuring a beam spin asymmetry (BSA). More accurately,
the difference of cross-sections, weighted by a calculable kinematical term directly relates to linear combinations of
GPDs. To some extent, it is still possible to measure the usual (normalized) BSA but only comparison with models are
possible. The presented data is not accurate enough to extract relevant GPD information, therefore all experiments using this
technique so far, have only published the result for their beam spin asymetry, along with model predictions.

\begin{figure}[hbt]
\begin{center}
\epsfig{file=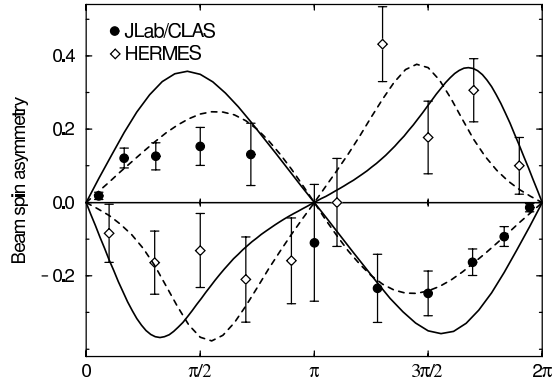,width=0.45\linewidth}
\end{center}
\caption{Dependence of the beam spin asymmetry as a function of the azimuthal angle for both the HERMES \cite{dvcs_hermes}
and CLAS experiments \cite{dvcs_clas}.
The opposite sign is explained by the opposite charge of the lepton beam used in both facilities. GPD models with different parameters
have been overlayed and show that theory and data are in fairly good agreement.}
\label{fig2}
\end{figure}

HERMES used its data collected with a 27.6~GeV positron beam impinging on an internal gaseous hydrogen target \cite{dvcs_hermes}.
The DVCS process was identified through the detection of the (e,$\gamma$) final state. The missing proton was checked
through the missing mass method. Due to the limited resolution of the HERMES spectrometer, the missing mass window around the
proton mass was $-1.5<M_X<1.7$~GeV. Exclusivity was checked through a Monte-Carlo, but is certainly not ensured and
parasitic asymmetries may arise for instance from nucleon resonances. In any case, HERMES obtained a sizeable BSA as shown on Figure~\ref{fig2}.
The amplitude of the $\sin\varphi$ term which is the GPD-linear term was measured to be -0.23 $\pm$0.04 (stat.) $\pm$0.03 (syst.)
at $<Q^2>=2.6$~GeV$^2$, $<x_{B}>=0.11$ and $<-t>=0.27$~GeV$^2$. Models, once again, show a reasonable agreement with data.

%

The same observable was measured at Jefferson Lab with a 4.2~GeV electron beam on a liquid hydrogen target \cite{dvcs_clas}.
Only the scattered
electron and recoil proton were detected in the Cebaf Large Acceptance Spectrometer (CLAS) installed in Hall~B. The outgoing
photon was identified through the missing mass method. The $\pi^0$ contamination was estimated bin-by-bin and subtracted using
a fitting method. The exclusivity was therefore ensured globally rather than event-by-event. The BSA shown on Fig.~\ref{fig2} 
was measured at $<Q^2>=1.25$~GeV$^2$, $<x_{B}>=0.19$ and $<-t>=0.19$~GeV$^2$. The amplitude of the $\sin\varphi$ term turned out to
be 0.202 $\pm$0.028 (stat.) $\pm$0.013 (syst.). The sign difference with respect to the HERMES result is easily explainable
since HERMES used a positron beam whereas the Jefferson Lab experiment used electrons. Figure~\ref{fig2} also shows a comparison
with models, which turn out to be in fair agreement considering that the theory predicted these curves much before data was ever shown.
Moreover, to get sufficient statistics, data is integrated over a rather large acceptance at this stage, which makes the comparison
with theory evaluated at the average kinematics point rather delicate.

Using 4.8~GeV and 6~GeV data at Jefferson Lab in Hall~B, an attempt was made to extract $Q^2$ and $t$ dependences \cite{dvcs_gagik,dvcs_harut}.
However, the data can only be compared to theory for now due to coarse statistical accuracy and large integration domains over
the remaining kinematical variables. In any case, they clearly show that binning
on all kinematic variables is possible with large acceptance detectors such as CLAS in Hall~B.


Whereas the DVCS Beam Spin Asymmetry accesses the DVCS amplitude's imaginary part, the Beam Charge Asymmetry (BCA) allows one to measure
the real part. Using both electron and positron 27.6~GeV beams, HERMES was able to extract a preliminary BCA \cite{dueren_contrib}.
Let us point out that this
measurement is very difficult since data sets separated by several years need to be subtracted to compute this BCA. Nevertheless,
HERMES proved it is a doable measurement and both the imaginary and real parts are needed to collect all the information about GPDs.

Additional analysis work is currently underway in all facilities to extract more
information from existing data. For instance, HERMES is currently working on kinematical dependence of its BSA observable,
with more statistics, as well as DVCS on nuclei. The Jefferson Lab CLAS collaboration has preliminary results concerning
Deuteron DVCS as well as Target Spin Asymmetries.

\section{NEW GENERATION OF DVCS EXPERIMENTS AT JEFFERSON LAB}

The results discussed in the previous section came within the last three years. Three dedicated DVCS experiments were scheduled to run in
2004-2005 at Jefferson Lab, two in Hall~A which have already taken data and one in Hall~B scheduled for the first half of 2005.

\subsection{Proton and neutron DVCS in Hall~A}

E00-110 and E03-106 in Hall~A are DVCS experiments at 5.75~GeV electron beam energy
(maximum possible energy at Jefferson Lab at the time). The former experiment is aimed at studying DVCS on the proton. As explained
in the introduction, it uses the feature that in photon electroproduction, BH and DVCS interfere and the resulting asymmetry is
linear in the GPDs. This experiment uses the Hall~A High Resolution Spectrometer (HRS) to detect the scattered electron, a lead-fluoride
calorimeter to detect the emitted photon and an array of out-of-plane scintillator blocks to detect the recoil proton. This experiment
requires the 3-particle final state to be entirely detected and is therefore the first truly exclusive DVCS experiment to date. An
important feature of this experiment is the high 10$^{37}$~cm$^{-2}$s$^{-1}$ luminosity and therefore a very accurate measurement
of the helicity dependent cross-sections. From this, both the cross-section difference and total cross-section can be obtained.
Three kinematic points have been investigated at fixed $x_B=0.35$: $Q^2$=1.5, 1.9 and 2.3~GeV$^2$. One of the goals of the experiment is
to check that the Bjorken regime has been obtained at this moderate energy and $Q^2$. This is done by examining
the $Q^2$ dependence of the properly weighted cross-section difference at fixed $x_B$ and compare it with its prediction.

\begin{figure}[hbt]
\begin{center}
\epsfig{file=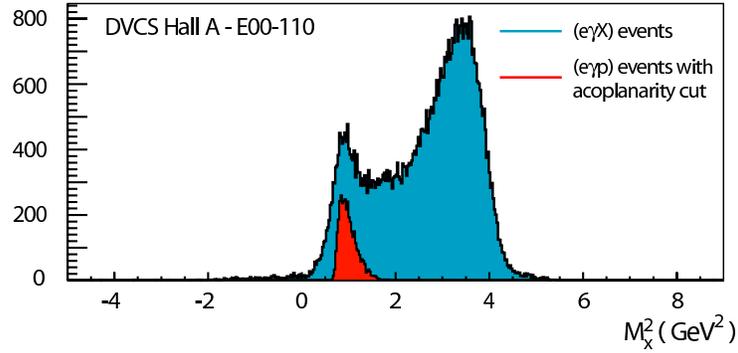,width=0.6\linewidth}
\end{center}
\caption{Missing mass squared using the detected electron and photon. The same distribution is plotted,
once a triple coincidence, both in time and space (acoplanarity), is required. This cut dramatically reduces
background events and ensures a high purity of exclusive DVCS events.}
\label{fig4}
\end{figure}

Since this experiment has already taken data, it is interesting to see the major improvement in requiring a 3-particle final state
with respect to the exclusivity. Figure~\ref{fig4} shows the missing mass squared for the Hall~A E00-110 experiment.
It is obvious that requiring all final state particles to be detected dramatically improves the missing mass spectrum
and gives much more confidence in the exclusivity of the reaction. Results are expected for the end of 2005.

%

Using the same base equipment, Hall~A E03-106 studied neutron DVCS using a liquid deuterium target as quasi-free neutrons. As was said
before, the properly weighted cross-section difference for opposite beam helicities is linear in the GPDs. At moderate $Q^2$, this
linear combination of GPDs is dominated by GPDs $H$ and $\widetilde{H}$ for a proton target. However, the linear combination
is dominated by the GPD $E$ for a neutron target due to the fact that the Pauli form factor is much larger than the Dirac form
factor in this case. Since the target is deuterium, both proton DVCS and neutron DVCS events are recorded. An additional veto detector
was installed in front of the scintillator-array to discriminate protons from neutrons in the final state.

\subsection{Proton DVCS in Hall~B}

\begin{figure}[hbt]
\begin{center}
\epsfig{file=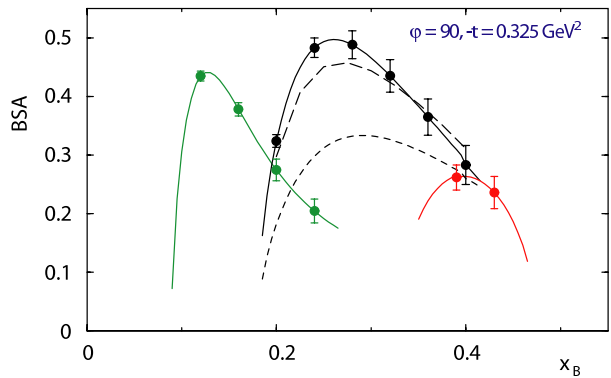,width=0.5\linewidth}
\end{center}
\caption{Simulated data for the BSA as a function of $x_B$ at fixed $\varphi$ and $t$, for 3 different $Q^2$ ranges, as expected
from Hall~B E01-113 experiment.}
\label{fig5}
\end{figure}

The DVCS experiment in Hall~B is scheduled to run in the first half of 2005 for about 2~months at a luminosity
of 5$\cdot$10$^{34}$~cm$^{-2}$s$^{-1}$ on a liquid hydrogen target. The CLAS spectrometer has been complemented by
an internal photon calorimeter at very forward angles. This calorimeter is composed of 424 lead-tungstate crystals
read-out by Avalanche Photodiodes. The calorimeter is protected from low-angle M\o{}ller electrons by a super-conducting
solenoid magnet with central field of about 5~T. A very successful test
run was achieved in 2004 to check the behavior of the calorimeter in such a high rate environment. Once again, this dedicated
experiment will require the full final state to be detected to suppress all competing channels and background. The
main difference with the previous Hall~A experiment is the large kinematical domain, allowing for the first time to check
kinematic dependences along the $x_B$ and $t$ variables independently, as shown on Fig.~\ref{fig5}.

%

\section{FUTURE DES MEASUREMENTS}

\begin{figure}[hbt]
\begin{center}
\epsfig{file=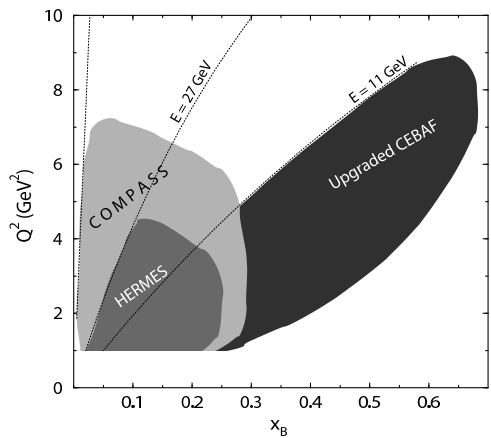,width=0.4\linewidth}
\end{center}
\caption{Ranges in $Q^2$ vs. $x_B$ for the COMPASS project, HERMES and CLAS after the Jefferson Lab 12~GeV upgrade.}
\label{fig6}
\end{figure}

It was obvious from the re-analysis of data described in section \ref{oldies}, that new experiments were necessary to extract
meaningful information about GPDs. The first three experiments will provide the first important steps in this direction with
much more statistics and true exclusivity. However, the GPDs depend on three kinematical variables and as such, require
a lot of data to have a complete view of the nucleon structure. Four experimental programs plan to make measurements of 
various electroproduction processes in the Bjorken regime in the near future: the H1 collaboration plans to use a recoil detector to tag
the recoil proton and therefore ensure the exclusivity of the reaction; HERMES will use the same strategy to improve their
first measurement; The COMPASS experiment at CERN with new detectors plans to make such measurement around 2010;
Jefferson Lab with upgraded energy up to ~12~GeV and new detection systems in all experimental Halls will make high statistics
measurement in a wide kinematical range. Figure~\ref{fig6} shows the $Q^2$ vs. $x_B$ range for the quark sector experiments. As for
H1, they will make measurements in the gluon sector at very low $x_B$ and high $Q^2$.

%

\subsection{H1 at HERA}

The H1 experiment at HERA recently added a Very Forward Proton Spectrometer (VFPS) in the proton beam line in order to efficiently
study diffractive events such as DVCS. This will allow to extend the measurements to lower values of $Q^2$ and $W$ with a
better $t$ measurement. Data taking will start again this year and will last until 2006 and perhaps beyond. 

\subsection{COMPASS}

\begin{figure}[hbt]
\begin{center}
\epsfig{file=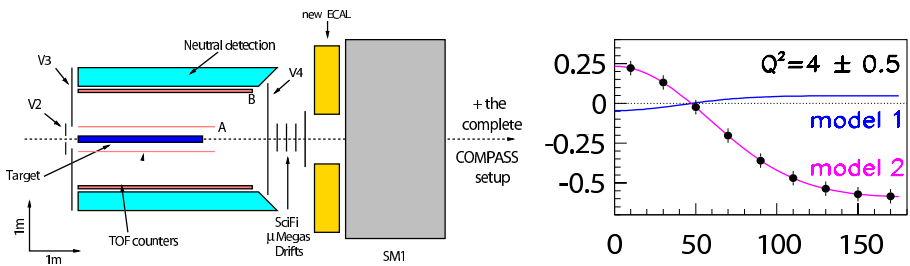,width=0.75\linewidth}
\end{center}
\caption{Left: Proposed setup for a DVCS measurement at COMPASS. A recoil detector, an extended calorimetry and a veto for charged forward
particles until 40 degrees have been added. Right: Projected error bars for a 6-month measurement of the azimuthal angular distribution of the BCA
at COMPASS for muon energy of 100~GeV. This particular bin is taken at $x_B=0.05\pm 0.02$ and $Q^2=4\pm 0.5$~GeV$^2$.}
\label{fig7}
\end{figure}

The COMPASS experiment at CERN is currently investigating the polarization of the gluons inside the nucleon. However, the collaboration
is discussing the possibility to perform a study of Deep Exclusive Scattering using an upgraded version of the detector \cite{compass_future}.
The high energy muon beam at CERN allows to measure both the hard exclusive meson production as well as DVCS
in the range $1.5\le Q^2\le 7.5$ and $0.03\le x_B\le 0.25$. The meson production mechanism can be investigated using the present
setup whereas the DVCS study requires the addition of a recoil proton detector, the extension of calorimetry to large angles for the
detection of the radiated photon and a veto for charged forward particles until 40$^\circ$ (Fig.~\ref{fig7} left).
In both cases, a 2.5~m-long liquid hydrogen target is necessary to provide a luminosity
of 5$\cdot$10$^{32}$cm$^{-2}$s$^{-1}$ which is required
to study these exclusive processes. The COMPASS DVCS experiment will be able to measure the total cross-section as well as the
Beam Spin and Beam Charge Asymmetries (Fig.~\ref{fig7} right) as a function of the azimuthal angle. The charge of the produced muon is
fully correlated with its polarization since the secondary muon beam comes from pion and kaon decay and provides a natural polarization.
After the proposal is accepted, the DVCS experiment at COMPASS could potentially start taking data in 2010.

%

\subsection{HERMES}

\begin{figure}[hbt]
\begin{center}
\epsfig{file=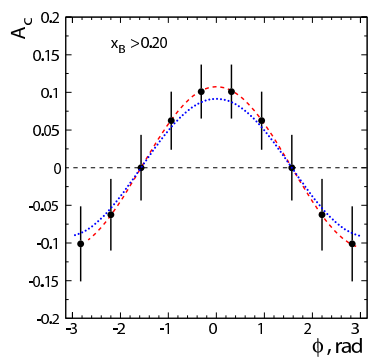,width=0.3\linewidth}
\end{center}
\caption{Expected accuracy for the BCA azimuthal distribution at HERMES during the HERA-II period, once the recoil proton detector is installed.}
\label{fig8}
\end{figure}

In order to meet the exclusivity requirement for DES processes, the HERMES collaboration will install a new recoil detector for
the envisaged final two years of the HERA-II running period. This new detector, which surrounds the internal gas target, focuses
on positive identification of recoiling protons, improving the transverse momentum reconstruction of these particles and rejecting
non-exclusive background events. It is planned to be installed in 2005. During the next two years of data taking, the collaboration
plans to measure the Beam Spin and Charge Asymmetries with fairly good accuracy, compared to their current measurements, as seen on
Fig.~\ref{fig8}.

%

\subsection{Jefferson Lab at 12~GeV}

\begin{figure}[hbt]
\begin{center}
\epsfig{file=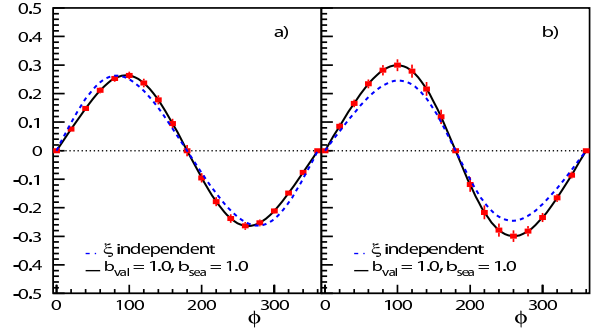,width=0.48\linewidth}
\end{center}
\caption{Expected data for the BSA at
$x_B=0.35$ and $-t=0.3$ (GeV/c)$^2$, and a)
$Q^2=2.75$ (GeV/c)$^2$ and b) $Q^2=5.4$ (GeV/c)$^2$. Data are simulated assuming 2000
hours of running at a luminosity of 10$^{35}$ cm$^{-2}$ sec$^{-1}$ with the upgraded CLAS detector.}
\label{fig9}
\end{figure}

In its pre-Conceptual Design Report, Jefferson Lab plans to use its upgraded 12~GeV facility to study Deep Exclusive Scattering processes
with an unprecedented statistical accuracy and in a large kinematical domain. Indeed, preliminary studies in all three Halls make DES one
of their top priority after the accelerator energy upgrade \cite{jlab_pcdr}. As an example, Hall~B plans to upgrade its CLAS spectrometer to
accomodate the energy increase: the torus magnet will be re-designed to take into account the change in momenta and trajectories
of scattered particles. The tracking system will be completly redesigned, whereas the calorimetry will just be complemented. Particle
identification, will be upgraded as well because of the change in momentum of the particles in the final state. Finally, a brand new central
detector composed of a solenoidal magnet with its associated detectors, will focus on the detection of particles emitted in the range
40-135$^\circ$. With this setup, CLAS++ will be able to handle luminosities as high as 10$^{35}$cm$^{-2}$s$^{-1}$, allowing for very
good statistics in a large kinematical range and within a reasonable timeframe. Thanks to the versatility of the CLAS++ detector,
all DES processes will be studied with this setup, and as an example, promising projected data for BSA can be seen on Fig.~\ref{fig9}.
Data taking is expected to begin as early as 2010.

%

\section{CONCLUSION}

Both the published work on DVCS from H1, ZEUS, HERMES and JLab and the new preliminary results are very promising. All these studies
demonstrate the feasability of larger-scale experiments aimed at studying DES processes in various kinematical domains. However,
the exclusivity and the statistical accuracy require new generations of experiments with dedicated setups and beam time.  Three
new DVCS experiments at Jefferson Lab in Halls A and B will give the first truly exclusive and statistically
significative results in 2005-2006. Following this, both H1 and HERMES will take data during the HERA-II run period with an upgraded
recoil detector. At the horizon of 2010, the COMPASS collaboration at CERN have started to investigate the possibility to make very
interesting measurements, especially of the BCA for DVCS. Finally, Jefferson Lab with its upgraded 12~GeV beam energy will cover a large
kinematical domain with high statistics and high resolution measurements of all DES processes.

The author wishes to thank N.~d'Hose, M.~Dueren and L.~Schoeffel for useful discussions about the DES experimental program at COMPASS, HERMES
and H1.

\end{document}